\documentclass[epj]{webofc}
\usepackage[varg]{txfonts}   % Web of Conferences font
\usepackage{graphicx,color} % for figures
\usepackage{bm}       % bold math 
\usepackage{amsmath}  
\usepackage{amssymb}
\woctitle{21st International Conference on Few-Body Problems in Physics}
\begin{document}
\title{Impressions of the Meson Spectrum:\\ Hybrids \& Exotics, present and future }
\author{M.R. Pennington\inst{1}\fnsep\thanks{\email{michaelp@jlab.org}}
% \and
%        Second author\inst{2}\fnsep\thanks{\email{secondauthor@alab.gov}} \and
%        Third author\inst{3}\fnsep\thanks{\email{Emailaddressforthirdauthorifnecessary}}
        % etc.
}

\institute{Theory Center, Thomas Jefferson National Accelerator Facility, 12000 Jefferson Avenue,\\ Newport News, VA 23606, U.S.A. 
%\and
%           the second here 
%\and
%           Last address
          }

\abstract{It has long been expected that the spectrum of hadrons in QCD would be far richer and extensive than experiment has so far revealed. While there have been experimental hints of this richness for some time, it is really only in the last few years that dramatic progress has been seen  in the exploration both experimentally and in calculations on the lattice. Precision studies enabled by new technology both with detectors and  high performance computations are converging on an understanding of the spectrum in strong coupling QCD. 
 These methodologies are laying the foundation for a decade of potential discovery that electro and photoproduction experiments  at Jefferson Lab, which when combined with key results on $B$ and charmonium decays from both $e^+e^-$ and $pp$ colliders, should turn mere impressions of the light meson spectrum into a high definition picture. 
}
\maketitle
\section{Introduction}
\label{intro}
The Chicago Art Institute is one of the great galleries of the world. It contains some of the best known Impressionist paintings
of the 19th century: Monet's series of ``waterlilies'' and ``haystacks''. This talk is also  an impression: an impression of the meson spectrum. The details may be fuzzy, but nevertheless there is a picture to be seen in a certain light. That light is the development of a detailed understanding of the workings of  strong coupling QCD.

The spectrum of hadrons is, of course, a direct reflection of the degrees of freedom that underly hadronic states, and the forces between these. For the ground state  baryon amd meson families, these appear to be just the 3 flavors of light quarks. For excited mesons, those with the maximum spin for a given mass, the $J^{PC}$ $= 1^{--},\, 2^{++},\, 3^{--}, \cdots$ {\it i.e.} the $\rho, \omega, K^*, \phi$,  $f_2(1270), a_2(1320), K_2^*(1430), f_2'(1525)$, $\cdots$ families, all appear strongly in hadro-production processes through their decays to $\pi\pi$, $K\pi$, 3$\pi$ and ${\overline K}K$. All belong to nearly ideally mixed quark model multiplets. These states are said to lie along the leading Regge trajectories.  They have set the paradigm for our idea of mesons. Their properties have defined the development of constituent quark models. Of course, QCD suggests that the hadronic states we observe are color singlets: the simplest being a quark and antiquark or three quarks. However QCD predicts there should be a far richer spectrum, with states made predominantly of glue, we call glueballs, tetraquark states made of two quarks and two antiquarks, some of which may like  to rearrange themselves for good dynamical reasons as hadronic molecules, or states in which glue contributes to their spatial quantum numbers  --- states we call hybrids, with analogs to be found in the baryon sector too.  For almost forty years we have been searching for these additional states. Indeed we may well have observed some of these, but there is little certainty of what has been found. They are in general more difficult to identify than the ideally mixed ${\overline q}q$ states. They have lower spins than the maximum at a given mass, and tend to be broad or only produced with smaller couplings, or both. Consequently one has to have data of high statistics with near perfect angular coverage to be able to perfom a robust Amplitude Analysis that can separate out the smaller partial waves with precision. 
It is only now with data  on hadroproduction from COMPASS@CERN, from BaBar@SLAC, Belle@KEK, BESIII@IHEP and LHCb@CERN, and shortly with GlueX and CLAS12@JLab, that we have (and will continue to have) such data.

Historically the focus for a richer spectrum was on the pure glue sector~\cite{Klempt:2007cp}. It is there where lattice calculations can most easily be done in a world with no quarks. These~\cite{{Sexton:1995kd},{Morningstar:1999rf},{Michael:1988jr}}  predicted the lightest glueball would be a scalar around 1.6-1.7~GeV, with a tensor higher up above 2 or 2.2 GeV. These were sought in what were thought to be glue-rich environments like $J/\psi$ radiative decays, or ${\overline p}p$ annihilation. Of course, glueballs decay to flavorless combinations of pions and kaons. This means they inevitably mix with ${\overline q}q$ states of the same quantum numbers that decay to the very same channels. Though there are indeed more isoscalar scalars, below 1.8~GeV, than can fit into two nonets, which (if any) is the closest to a glueball is matter of debate and modeling~\cite{{Close:2000yk},{Ochs:2013gi},{Janowski:2014ppa}}. Attention consequently turned to multi-quark states, some of which might indeed be the light scalar mesons. However, it is in the heavy flavor sectors of charmonium and bottomonium that mesons that are more than just ${\overline q}q$ have been most spectacularly seen.    
\begin{figure}[b]
\vspace{0.3cm}
\centering
\sidecaption
\includegraphics[width=9.cm,clip]{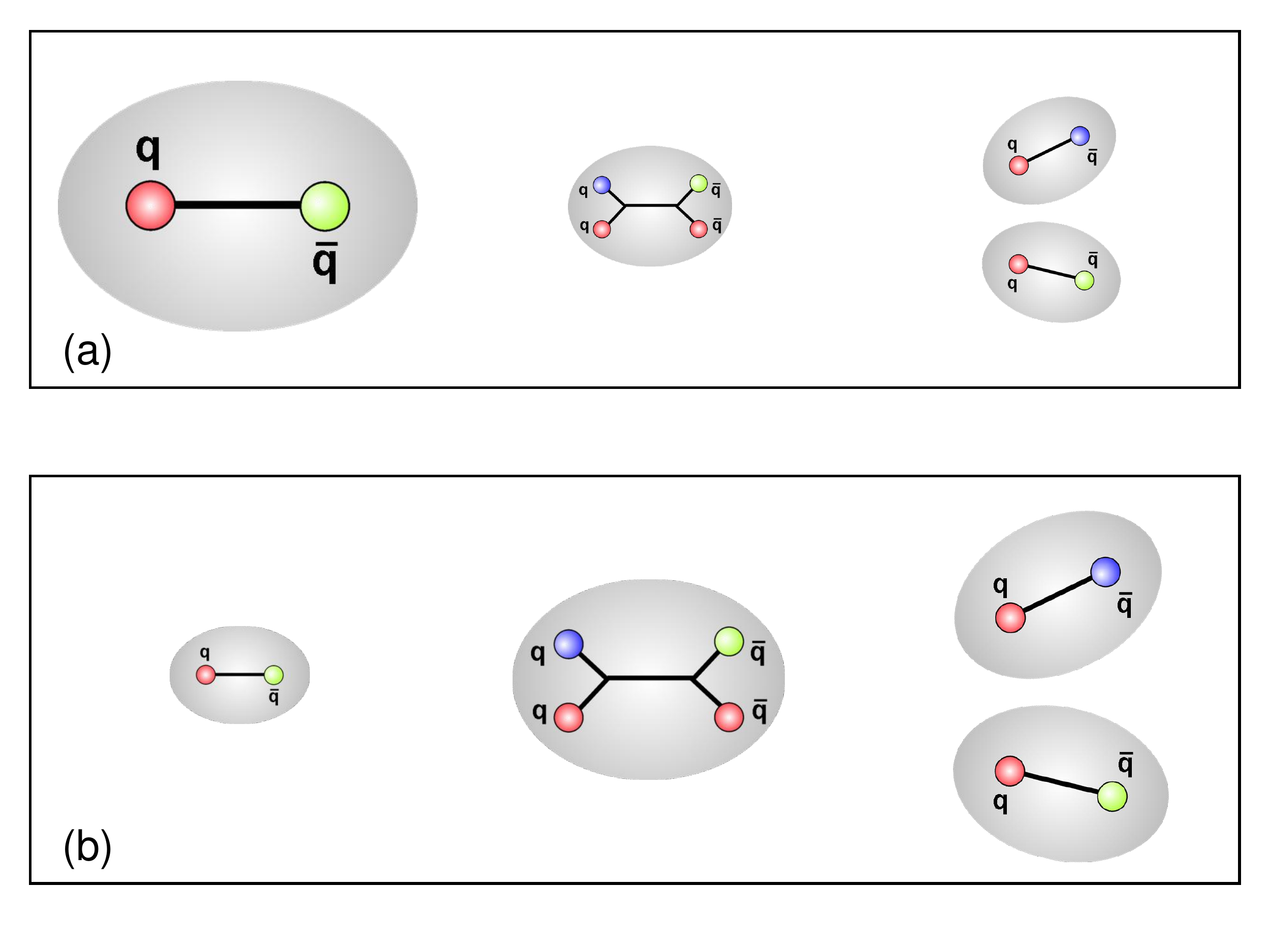}
\caption{Cartoon of the Fock space of two types of mesons. In the upper graph (a) is shown a predominantly ${\overline q}q$ meson with smaller four quark and di-meson components, and in the lower (b) a possible orthogonal state, which though {\it seeded} by ${\overline q}q$ configurations, may be dominated by tetraquark or di-hadron components~\cite{{vanBeveren:1986ea},lutz,{Pennington:2014tpa}}.}
\label{fig-1}       % Give a unique label
\end{figure}

\section{The Richness of the Hadron Spectrum} 
While the very lightest mesons and baryons are stable as far as  isospin conserving strong interactions are concerned, all excited states decay. Indeed, they decay very quickly. While they may be seeded by configurations with the minimum number of quarks, they almost immediately radiate gluons that create additional ${\overline q}q$ pairs. It is through these that the decay products emerge. Decays are an integral part of the lifecycle of all excited hadrons. The $\rho^+$ is not simply $u{\overline d}$, but composed of additional $u{\overline u}$ and  $d{\overline d}$ pairs. These rearrange themselves into a $\pi^+\pi^0$ final state. Thus as illustrated in Fig.~1a, the Fock space of the $\rho^+$ is not simply $u{\overline d}$. Nevertheless, this component dominates: the coupling to the $\pi\pi$ system is inhibited by being a $P$-wave interaction. Indeed all the states on the leading Regge trajectory, the easiest to be seen in experiment, are dominated by their $q{\overline q}$ configurations. However, lower lying states, like the chiral partner of the $\rho$, the $a_1(1260)$, spends more of its time in a multi-quark configuration, first $\rho\pi$ and then 3$\pi$. Ideal mixing is less apparent. As soon as one recognizes that the degrees of freedom are not just ${\overline q}q$, but include multiquark or multihadron configurations too, one understands that if the dynamics is favorable, there should also exist orthogonal states, in which the ${\overline q}q$ {\it seed} might be small and the dihadron components  large, as illustrated in Fig.~1b. 
Just such a possibility appears to occur with the $X, Y, Z$ states (or at least some of them). The opening of decay channels of the heavier charmonium states shifts them  from where they would naively be in a potential model, and at the same time new states appear, Fig.~1. The longest known is the $X(3872)$ seen in $B$-decay to $KX$, where $X \to J/\psi \pi\pi$. While this state is 140 MeV above ${\overline D}D$ threshold, it lives 100 times longer than expected (with a width of less than 1.2~MeV). It sits within a ``whisker'' of ${\overline {D^0}}D^{*0}$ threshold and 8~MeV below the corresponding charged $D$ threshold. This dynamics may make it favorable that one pion exchange binds the system~\cite{{Tornqvist:2004qy}}. Indeed, it may well be an example of a state with the Fock space shown in Fig.~1b --- see, for instance~\cite{Coito:2012vf}. 
\begin{figure}
\centering
%\sidecaption
\includegraphics[width=12.cm]{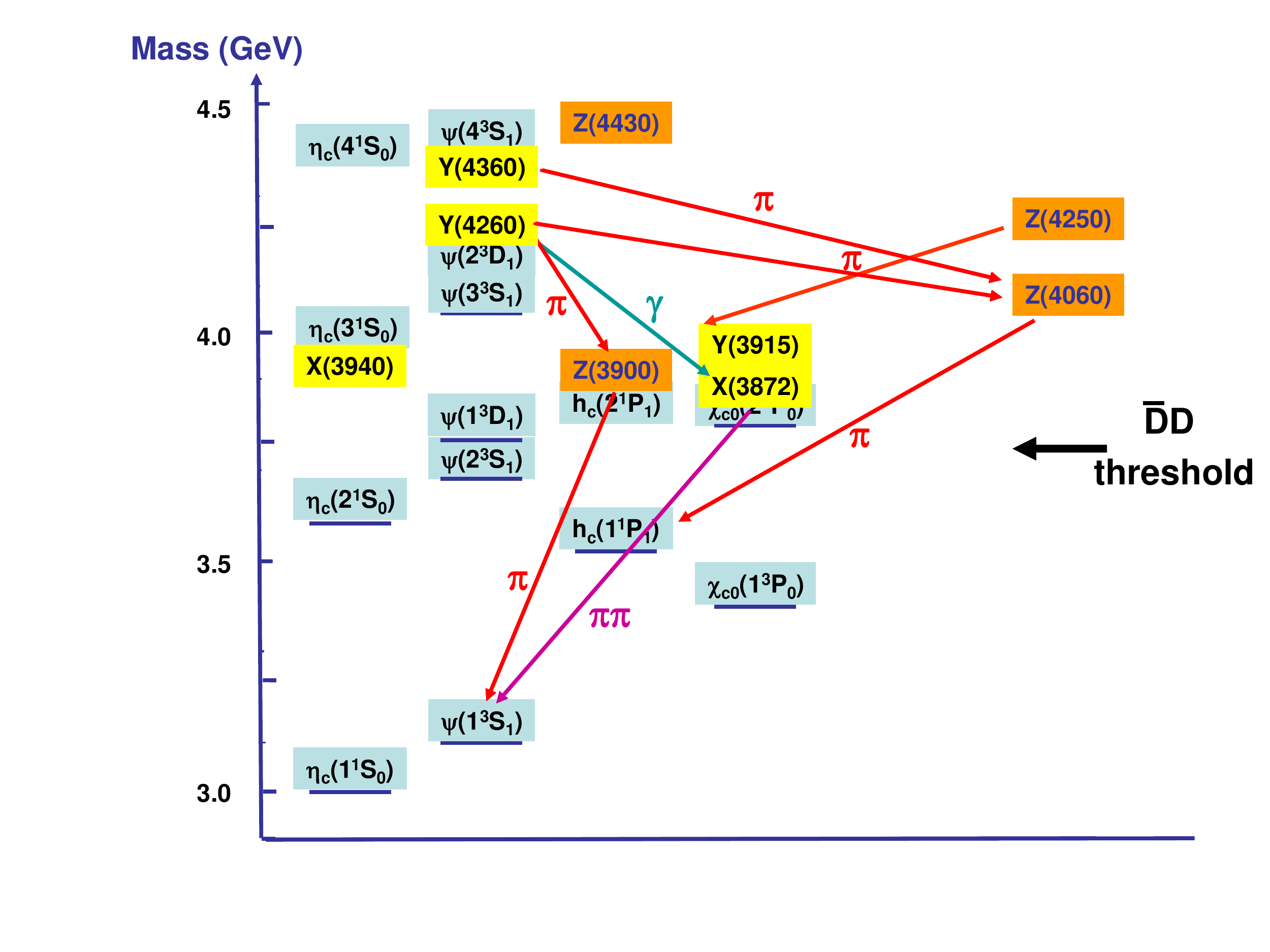}
\vspace{-0.5cm}
\caption{The standard charmonium spectrum with states (in light blue) labeled by their $n ^{2S+1}L_J$ quantum numbers, overlaid with the states of the {\it new} $X, Y, Z$ spectroscopy, showing some of their interconnections implied by their observed decay patterns~\cite{{Mitchell:2014zfa}}. }
\label{fig-2}       % Give a unique label
\vspace{-0.2cm}
\end{figure}

Since the discovery of the $X(3872)$, a whole host of these unusual mesons have been found. Not every bump need be due to a state in the spectrum. Consequently the new \lq\lq discoveries'' have really to be shown to give rise to poles in the complex energy plane, which is the true confirmation of a state in the spectrum. Nevertheless, there are too many to be accidental, Fig.~2. All may well be multi-quark states sitting close to a nearby open hadronic channel to which they couple in an $S$-wave. They may then be hadronic molecules or tetraquarks, Fig.~3. Indeed the appearance of the charged $Z_C(4430)$ points to just such configurations, definitely something more than ${\overline c}c$ as in Fig.~3. Two remarkable things hint to a relation between all these many states that Cheng-Ping Shen~\cite{shen} and Eric Braaten~\cite{braaten} will discuss in much more detail. First they seem to be connected to each other (Fig.~2), one decays into another by emitting a photon or a pion, depending on quantum numbers. The second thing is that there are similarly narrow states in the bottomonium sector too.   We don't yet have a working theoretical understanding of {\it all} of these states, though for 
some~\cite{{Braaten:2014qka},{Braaten:2013boa}}, as Eric Braaten will discuss, this may be closer. Nevertheless, when we do understand why and how such systems bind, this will surely teach us about the workings of strong coupling QCD. There are even hints that similar states may occur in the hidden strange sector with a $\phi f_0(980)$ bound system~\cite{Y(2175)}. Searches for ${\overline K} K^*$ systems, akin to ${\overline D}D^*$ of the $X(3872)$ will be a natural target for the future.
\begin{figure}
\centering
\sidecaption
\includegraphics[width=9.cm]{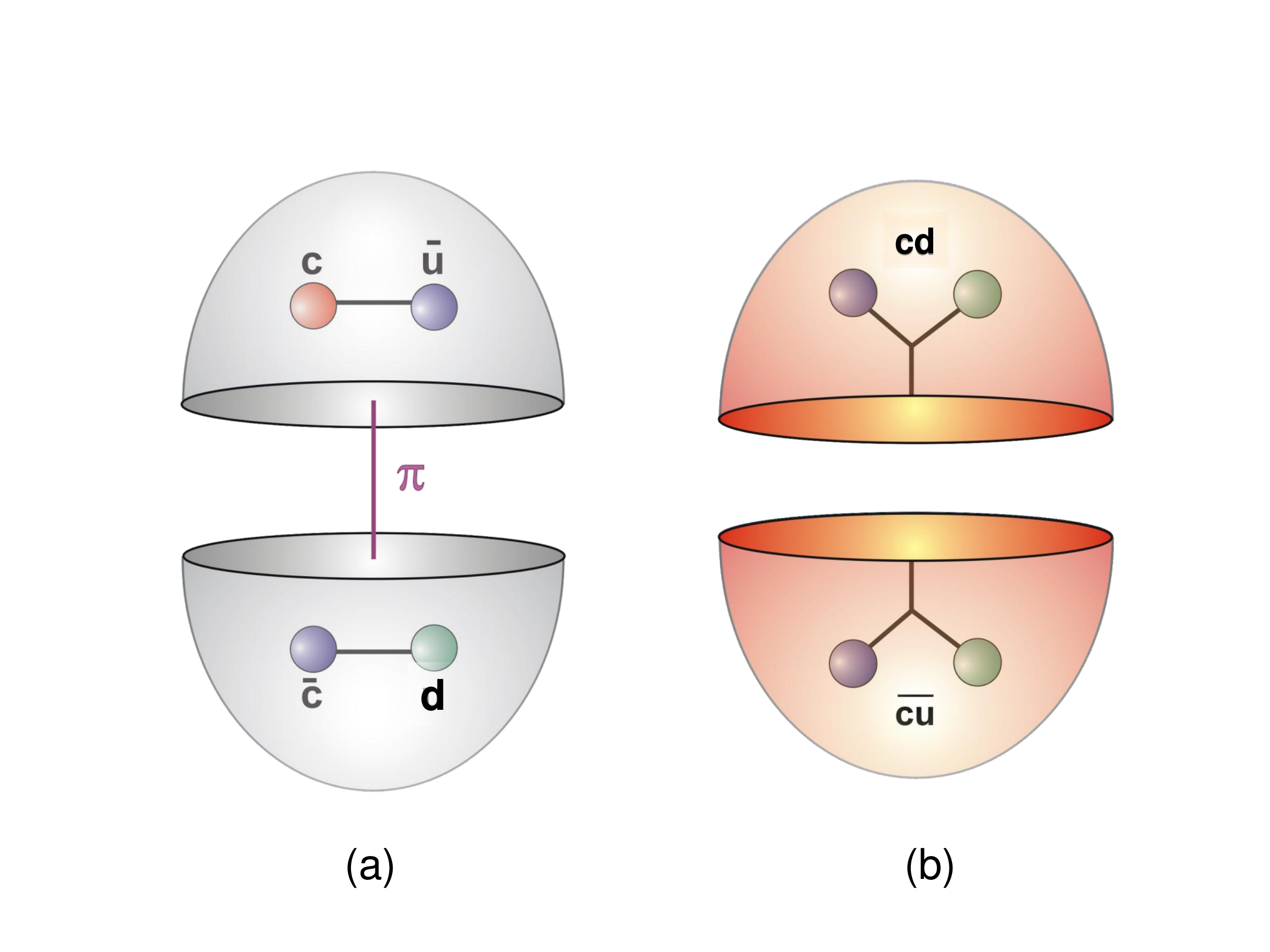}
\vspace{-0.5cm}
\caption{Representation of the $Z_C^-(4430)$ (a) as a molecule bound by interhadron forces, or (b) as a tetraquark state bound by interquark forces.}
\label{fig-2}       % Give a unique label
\vspace{-0.2cm}
\end{figure}

Indeed, it could well be that the nine lightest scalars, the $\sigma/f_0(500)$, $\kappa/K_0^*(800)$, $a_0(980)$ and $f_0(980)$, are of a similar type, tetraquark configurations, seeded by the heavier $0^{++}$ ${\overline q}q$ nonet underneath the corresponding $2^{++}$ states, but dominated in their Fock space by their di-meson components, as in Fig,~1b. In many respects~\cite{Pennington:2014tpa}, the $f_0(980)$ behaves as a ${\overline K}K$ molecule, probably the $a_0$ too (with the $a_0^-$ being like Fig.~3b with $s$ quarks replacing $c$'s), while the $\sigma$ and $\kappa$ are loosely bound $\pi\pi$ and $K\pi$ systems, respectively. If so, any simple quark mixing scheme for them is quite misleading~\cite{{Ochs:2013gi},{Fariborz:2006ff},{Black:2000nc}}. \lq\lq Mixing'' depends not just on the complex masses of the bound states, but on the masses of the pseudoscalar pairs to which they couple, and this has little (or nothing) to do with the intrinsic ${\overline q}q$,
or even glue, they each likely have in their Fock space.

\begin{figure}[t]
%\vspace{-3mm}
\centering
%\sidecaption
\includegraphics[width=12.5cm,clip]{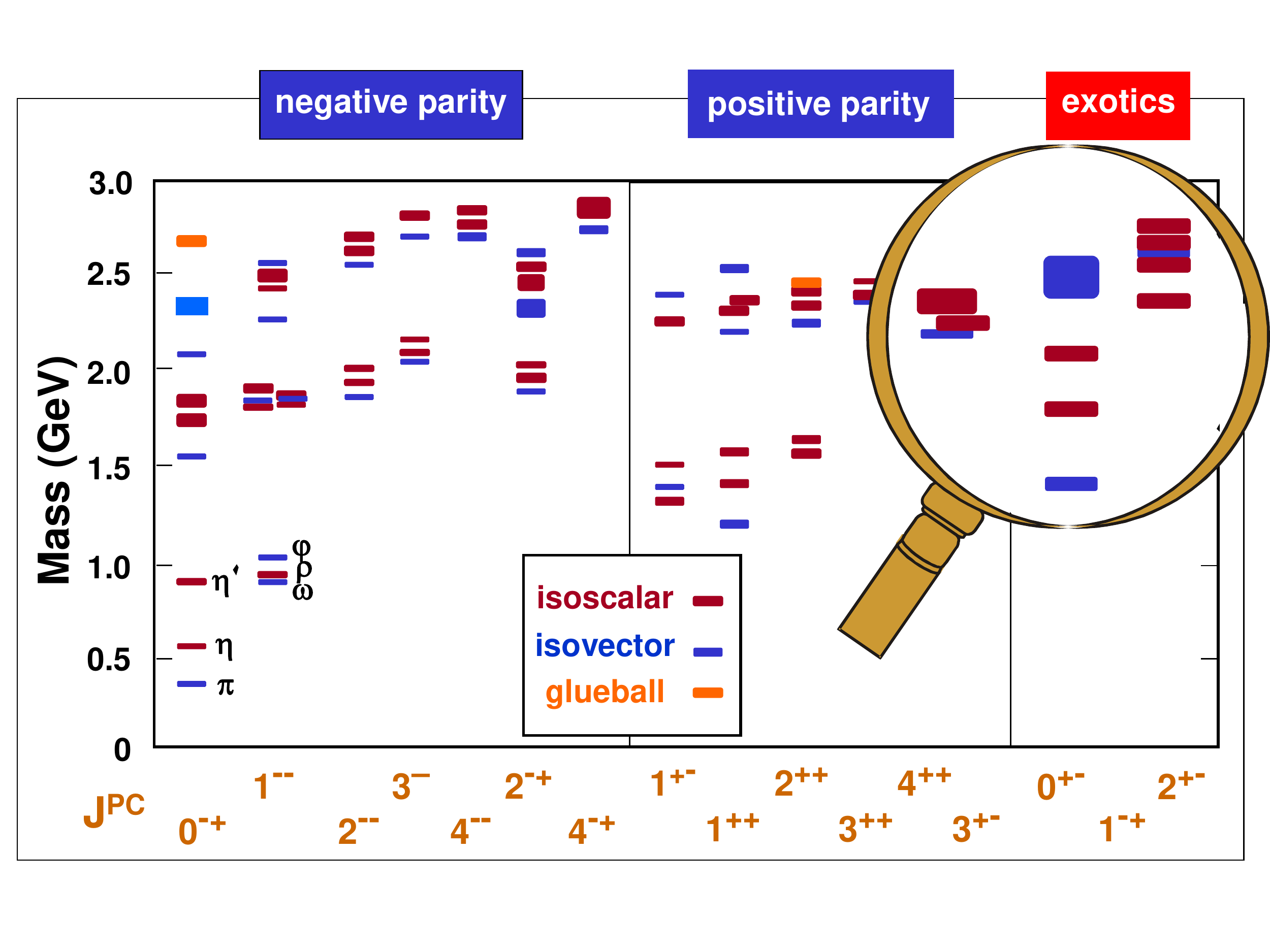}
\vspace{-0.8cm}
\caption{The masses  of the negative and positive parity mesons generated in QCD by calculating quark-antiquark operators on the lattice at a pion mass of 396~MeV, in columns labeled by their $J^{PC}$ values from the Hadron Spectrum Collaboration~\cite{{Dudek:2011tt},{Dudek:2009qf}}. The inclusion of gluonic operators adds hybrids with conventional quantum numbers on the left, roughly 1~GeV heavier than those with just quark operators, as well as generating states with \lq\lq exotic'' quantum numbers that are shown magnified on the right around 2~GeV in mass.}
\label{fig-4}       % Give a unique label
\end{figure}

Results from lattice QCD are having a direct impact on experimental studies now underway. The Hadron Spectrum Collaboration~\cite{Dudek:2009qf} several years ago computed the spectrum of mesons built from quark-antiquark operators with both isovector and isoscalar quantum numbers at a pion mass of 396~MeV. As seen in Fig.~4, this accords well with the observed spectrum, just shifted by having a heavy pion~\cite{{Dudek:2009qf},{Dudek:2011tt}}. Including gluonic operators enlarges the spectrum to include ${\overline q}gq$ hybrids. As well as producing additional (but heavier) mesons with \lq\lq conventional'' quantum numbers on the left in Fig.~4, it also creates states  (seen on the right) which are \lq\lq exotic'' in the quark model. The lightest of these hybrids, known as the $\pi_1$, has $J^{PC} = 1^{-+}$ in the mass range around 2~GeV, accessible to the 12~GeV  campaign at Jefferson Lab~\cite{opportunities}. Once such calculations include physical decay channels, with 140~MeV pions, the masses of the mesons are expected to be shifted, generally lower. To be certain that any state discovered is really a gluonic hybrid, it is essential to identify several such hybrid states with different flavor and spatial quantum numbers --- ideally whole multiplets.  This is what the GlueX experiment at JLab will put under the magnifying glass, Fig.~4. To~aid this further, the aim 
over the next few years is to predict using lattice QCD the decays and radiative transition amplitude of such hybrids. This follows the recent successful  benchmarking of such quantities for well-known mesons like the $\rho$, calculating its decay to $\pi\pi$ and its radiative transition from a pion~\cite{{Dudek:2014qha},{Wilson:2014cna},{Shultz:2015pf}, {Briceno:2015dca}}.

\begin{figure}[t]
\centering
\sidecaption
\includegraphics[width=5.cm,clip]{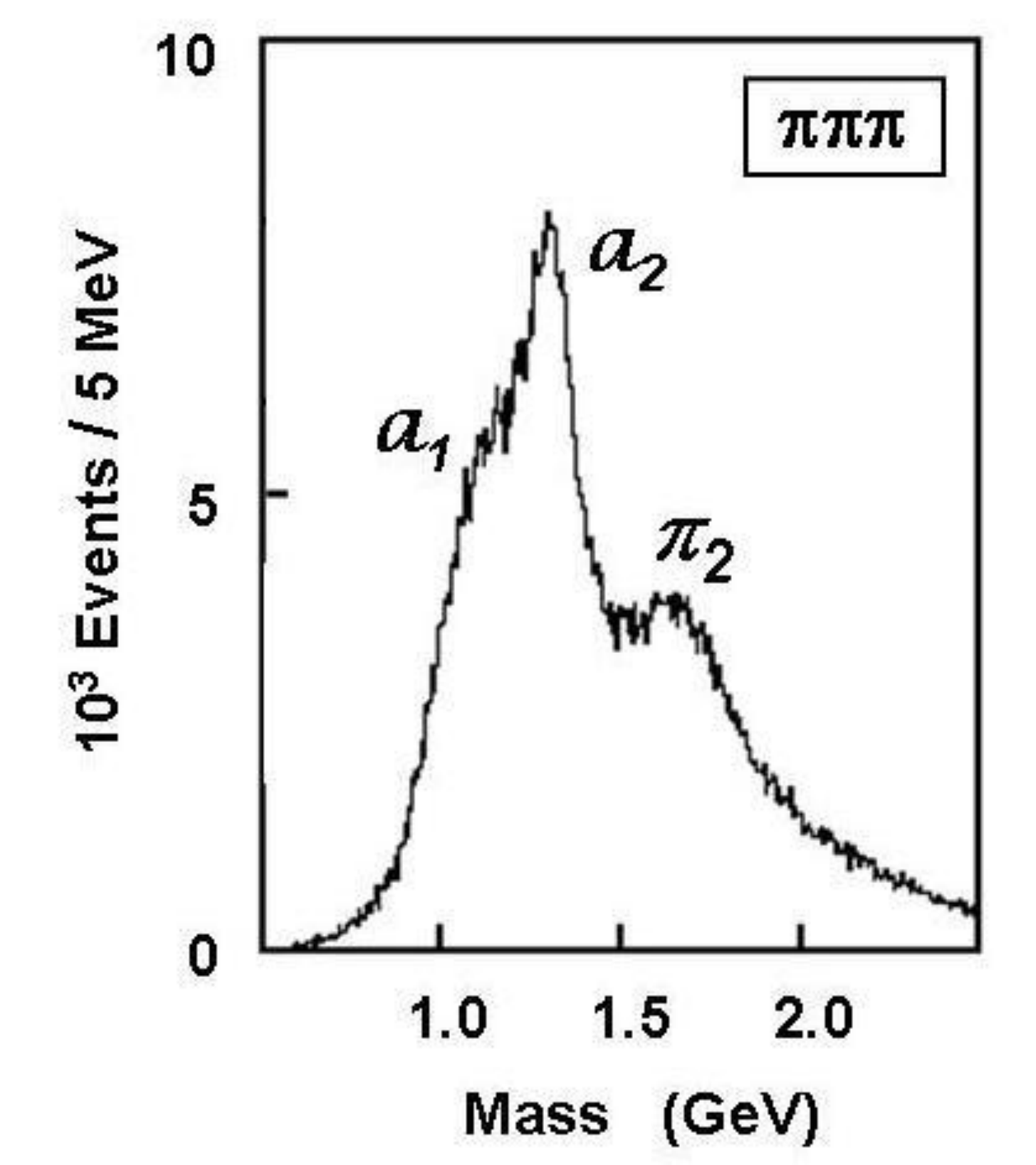}
\caption{The $\pi^+\pi^-\pi^-$ mass spectrum found in $\pi^-p$ collisions at 18 GeV/c by BNL-852 experiment~\cite{{Adams:2000zg}}. The dominant resonances are indicated. The COMPASS experiment at 190 GeV/c has 400 times the statistics on the same channel with a very similar, but inevitably smoother, 3$\pi$ spectrum~\cite{Krinner:2014mla}.}
%\vspace{-0.5cm}
\label{fig-8}       % Give a unique label
\end{figure}

Previously experiments at CERN and Serpukhov with the GAMS detectors~\cite{{Alde:1988bv},{Alde:1988mi}}, at Brookhaven with BNL-E852~\cite{{Adams:2000zg},chung} and later at Protvino with VES~\cite{{Beladidze:1993km},{Zaitsev:2000rc}} were aimed at finding states
with gluonic components. Indeed, they each found hints of mesons with $1^{-+}$ quantum numbers in $\pi\eta$ and $\pi\eta'$, but one could never be certain these were really resonances, rather than simple enhancements. It is in the $3\pi$ channel where larger statistics have been accumulated. So for instance in Fig.~5, we show the $\pi^+\pi^-\pi^-$ mass spectrum in $\pi^-p$ collisions at 18~GeV/c from BNL-E852~\cite{{Adams:2000zg}}. There the  high lying states, the $a_2(1320)$ and $\pi_2(1670)$, are easily seen. However, the much smaller $1^{-+}$ $\rho\pi$ $P$-wave signal was found to be highly sensitive to leakage from the higher waves, and to treatment of the large $\pi\pi$ $S$-wave interaction, and so claims of a $\pi_1(1400)$ came and went to be replaced by a possible $\pi_1(1600)$~\cite{chung,{Dzierba:2005jg},{Beladidze:1993km},{Zaitsev:2000rc},{Alekseev:2009aa}}. To prove the magnitude and relative phase variations arise from a pole in the complex energy plane requires more events with a hermetic detector and more detailed analysis.

\section{Challenges for the near future}
An attempt to use high statistics to completely clean up the light hadron spectrum is provided by combining results from several different production mechanisms to which new technology  allows precision access. Hadroproduction with pion beams at COMPASS, heavy flavor decays from BaBar, Belle, BESIII and LHCb, and photoproduction (with real and virtual photons) at Jefferson Lab with GlueX and CLAS12. Here is where a whole suite of precision analysis tools are required to match the precision of the existing data, or that to come, if we are to truly extract definitive physics results from these data.
COMPASS is the experiment at 190~GeV/c that provides the world's largest sample of three pion production with almost 100 Million events, four hundred times that of BNL-E852. The dominant partial waves (magnitudes and relative phases), which encode the $a_2(1320)$ and $\pi_2(1670)$, are robustly determined. Indeed, with such statistics, not only is the $a_1(1260)$ readily seen, as in Fig~4 in its $\rho\pi$ mode, but a new $a_1(1420)$ structure observed in the  hundred times smaller $f_0(980)\pi$ channel~\cite{{Adolph:2015pws}} -- see also~\cite{{Basdevant:2015wma}}. It is at this level that waves with \lq\lq exotic'' quantum numbers appear: they are only a tiny fraction of the integrated cross-section.
\begin{figure}
%\vspace{0.3cm}
\centering
%\sidecaption
\includegraphics[width=11.cm,clip]{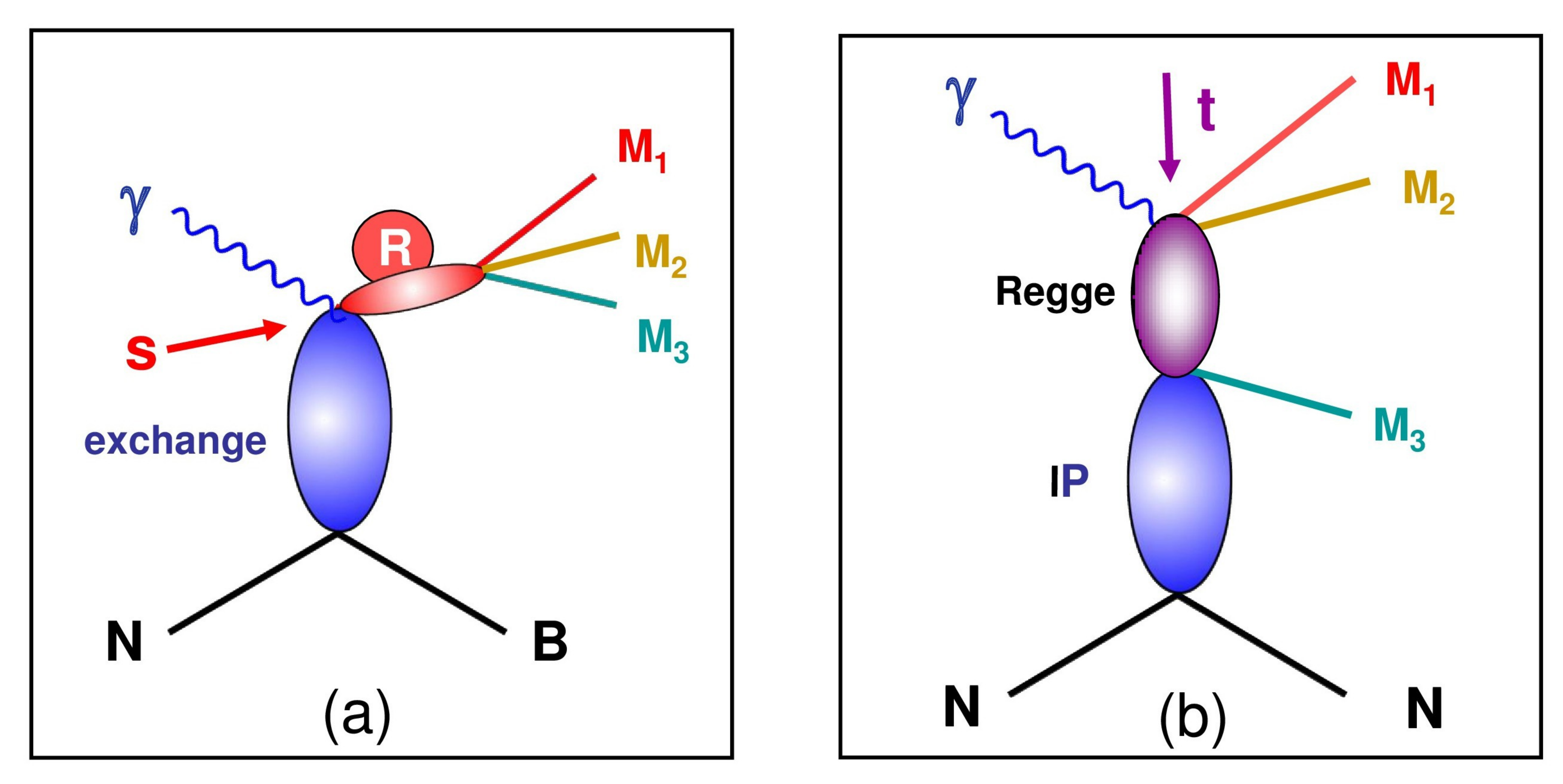}
%\vspace{0.4cm}
\caption{Examples of reaction mechanisms for multi-particle production. On the left (a) shows the peripheral mechanism for high energy photo-production of a state, with quantum numbers denoted by the circled $R$, which then decays to meson 1-3. This process is controlled by Pomeron and/or other Regge exchanges. The sum over channels of many quantum numbers $R$ (circled) in $\gamma$-Reggeon scattering is expected to have a simpler Regge exchange description, illustrated on the right (b).
Such a description provides a more economical representation of the higher waves than simply summing those on the left. Fig.~(b) also includes the case when the upper Regge exchange and meson $M_3$ are pions. This is commonly called the {\it Deck mechanism}~\cite{Deck:1964hm}.}
\label{fig-6}       % Give a unique label
\vspace{-0.6cm}
\end{figure}

 To extract partial waves which contribute less than a few percent, one needs to be certain that one understands all of the other waves to even better accuracy. This is a major challenge. Small partial waves are most readily affected by artificially limiting the number of  waves contributing. Something one had to do with earlier lower statistics data. COMPASS analyses~\cite{haas} have  dramatically illustrated the effect on the $1^{-+}$ $\rho\pi$ $P$-wave of different truncations  by showing how fitting their data with restricted wavesets reproduces the results of analyses by  BNL-E865 and by the VES group, matching the number of partial waves included by the BNL and VES experiments. The possible $\pi_1(1400)$ and $\pi_1(1600)$ broaden and change shape, as the number of contributing waves is altered~\cite{haas}.  To ensure this is not an issue, COMPASS includes a far bigger set of waves: 88 in total, a set that is deliberately  ``over-complete''~\cite{Krinner:2014mla}.

In all these studies, the right theoretical foundations are essential. The Joint Physics Analysis Center ({\it JPAC})  project~\cite{jpac} at Jefferson Lab has taken on the challenge to coordinate these devlopments. The S-Matrix has been reloaded.  Basic concepts of analyticity, crossing and unitarity well known for over 60 years provide the foundation for precision analysis of precison data. These are coupled with studies of the relevant reaction mechanisms, seen for instance in Fig.~6a, for hadro- and photo-production. These are key to avoiding the artificial truncation of partial waves. Cross-channel exchanges, seen in Fig.~6b, can provide a more efficient and effective way~\cite{{Deck:1964hm},{Bowler:1975my},{Basdevant:1977ya}}
 of handling higher partial wave amplitudes. Similarly, we know precision data demand more than a crude isobar picture of pairwise interactions. Yes, the $a_1$ (and even a $\pi_1$) may decay to $\rho\pi$ and then the $\rho$ decay to $\pi\pi$, but rescattering involving all three final state pions, Fig.~7, has to be taken into account to preserve unitarity. This is achieved employing the Khuri-Treiman equation~\cite{{Khuri:1960zz}}. {\it JPAC} calculations of  the analogous graphs for the $\eta$, $\omega$ and $\phi$~\cite{{Danilkin:2014cra},{Guo:2014vya},{Guo:2015zqa}} where high statistics data already exist,  reveal rescattering and 3-body interactions to be small for these decays. This is not surprising given the limited phase space. Nevertheless this benchmarking is essential before tackling heavier states, like the $a_1$'s and $\pi_1$, where such unitarity constraints become more complex as the number and energy range of waves in each interaction increases. Almost for the first time, data demand this level of detail.  
\begin{figure}[b]
%\vspace{-0.2cm}
\centering
%\sidecaption
\includegraphics[width=14.cm,clip]{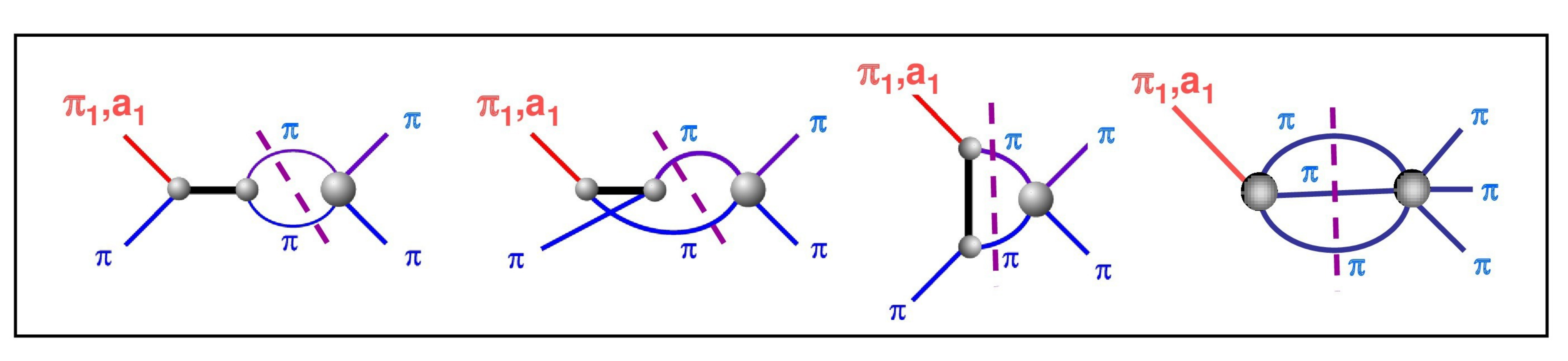}
%\vspace{-2cm}
\caption{An example of the three pion decay of a parent hadron, like the $a_1$ or a putative $\pi_1$. This has contribution from the pairwise interaction of the pions, and their rescattering, as well as the contribution of direct three body decay. All are required to ensure probability is conserved. The dashed lines indicate the 2 and 3-body contributions to the  unitarity relation.  Recent calculations, using the Khuri-Treiman equation~\cite{{Khuri:1960zz}}, of the analogous graphs for the $\eta$, $\omega$ and $\phi$~\cite{{Danilkin:2014cra},{Guo:2014vya},{Guo:2015zqa}} reveal rescattering and 3-body interactions to be small, because of the limited phase space for interactions any more complex than low energy $\pi\pi$ in an $S$ or $P$-wave. Calculations for the heavier states shown are underway. }
\label{fig-7}       % Give a unique label
\end{figure}

The richest source of three body decays is from $B$ and charm factories, from BaBar, Belle and BES, with rapidly appearing additions from LHCb. Again these are the channels with the largest statistics required to determine the closure of the weak unitarity triangle. The largest uncertainties in the CKM matrix parameters come from strong interaction effects from such multi-body decays. Thus the same technology of analysis required for understanding the hadron spectrum can inform and be informed by studying data from these machines. Cooperation and collaboration of theory and experiment, with data from different facilities is essential to progress.  It is only by such planned coordination that we can turn impressions of the meson spectrum, of hybrids and exotics, into a definitive picture that translates the confinement dynamics of strong coupling QCD into the detailed pattern of  hadrons we observe. That is the challenge for the next decade.

\begin{acknowledgement}
It is a pleasure to thank the conference organizers, particularly Charlotte Elster, Daniel Phillips and Craig Roberts, for inviting me to share my impressions of the meson spectrum.
This material is based upon work supported by the U.S. Department of Energy, Office of Science, Office of Nuclear Physics under contract DE-AC05-06OR23177."
\end{acknowledgement}

%
% BibTeX or Biber users please use (the style is already called in the class, ensure that the "woc.bst" style is in your local directory)
% \bibliography{name or your bibliography database}

\begin{thebibliography}{}
%\cite{Klempt:2007cp}
\bibitem{Klempt:2007cp} 
  E.~Klempt and A.~Zaitsev,
  %``Glueballs, Hybrids, Multiquarks. Experimental facts versus QCD inspired concepts,''
  Phys.\ Rept.\  {\bf 454}, 1 (2007)
  [arXiv:0708.4016 [hep-ph]].
  %%CITATION = ARXIV:0708.4016;%%
%\cite{Sexton:1995kd}
\bibitem{Sexton:1995kd} 
  J.~Sexton, A.~Vaccarino and D.~Weingarten,
  %``Numerical evidence for the observation of a scalar glueball,''
  Phys.\ Rev.\ Lett.\  {\bf 75}, 4563 (1995)
  [hep-lat/9510022].
  %%CITATION = HEP-LAT/9510022;%%
%\cite{Morningstar:1999rf}
\bibitem{Morningstar:1999rf} 
  C.~J.~Morningstar and M.~J.~Peardon,
  %``The Glueball spectrum from an anisotropic lattice study,''
  Phys.\ Rev.\ D {\bf 60}, 034509 (1999)
  [hep-lat/9901004].
%\cite{Michael:1988jr}
\bibitem{Michael:1988jr} 
  C.~Michael and M.~Teper,
  %``The Glueball Spectrum in SU(3),''
  Nucl.\ Phys.\ B {\bf 314}, 347 (1989).
  %%CITATION = NUPHA,B314,347;%%


%\cite{Close:2000yk}
\bibitem{Close:2000yk} 
  F.~E.~Close and A.~Kirk,
  %``The Mixing of the f(0)(1370), f(0)(1500) and f(0)(1710) and the search for the scalar glueball,''
  Phys.\ Lett.\ B {\bf 483}, 345 (2000)
  [hep-ph/0004241].
  %%CITATION = HEP-PH/0004241;%%
%\cite{Ochs:2013gi}
\bibitem{Ochs:2013gi} 
  W.~Ochs,
  %``The Status of Glueballs,''
  J.\ Phys.\ G {\bf 40}, 043001 (2013)
  [arXiv:1301.5183 [hep-ph]].
  %%CITATION = ARXIV:1301.5183;%%
%\cite{Janowski:2014ppa}
\bibitem{Janowski:2014ppa} 
  S.~Janowski, F.~Giacosa and D.~H.~Rischke,
  %``Is f0(1710) a glueball?,''
  Phys.\ Rev.\ D {\bf 90}, no. 11, 114005 (2014)
  [arXiv:1408.4921 [hep-ph]].
  %%CITATION = ARXIV:1408.4921;%%
%\cite{vanBeveren:1986ea}
\bibitem{vanBeveren:1986ea} 
  E.~van Beveren, T.~A.~Rijken, K.~Metzger, C.~Dullemond, G.~Rupp and J.~E.~Ribeiro,
  %``A Low Lying Scalar Meson Nonet in a Unitarized Meson Model,''
  Z.\ Phys.\ C {\bf 30}, 615 (1986)
  [arXiv:0710.4067 [hep-ph]].
  %%CITATION = ARXIV:0710.4067;%%
\bibitem{lutz} M.F.M.\ Lutz and E.E.\ Kolomeitsev, {\it Nucl.\ Phys.\ A} {\bf 755}, 29 (2005).
%\cite{Pennington:2014tpa}
\bibitem{Pennington:2014tpa} 
  M.~R.~Pennington,
  %``The Brief Life of a Hadron: QCD unquenched,''
  Acta Phys.\ Polon.\ Supp.\  {\bf 8}, 9 (2015)
  [arXiv:1411.7902 [nucl-th]].
  %%CITATION = ARXIV:1411.7902;%%
%\cite{Mitchell:2014zfa}
\bibitem{Mitchell:2014zfa} 
  R.~E.~Mitchell,
  %``Connecting the XYZ at BESIII,''
  PoS Hadron {\bf 2013}, 038 (2013).
  %%CITATION = POSCI,Hadron2013,038;%%
%\cite{Tornqvist:2004qy}
\bibitem{Tornqvist:2004qy} 
  N.~A.~Tornqvist,
  %``Isospin breaking of the narrow charmonium state of Belle at 3872-MeV as a deuson,''
  Phys.\ Lett.\ B {\bf 590}, 209 (2004)
  [hep-ph/0402237].
  %%CITATION = HEP-PH/0402237;%%
%\cite{Coito:2012vf}
\bibitem{Coito:2012vf} 
  S.~Coito, G.~Rupp and E.~van Beveren,
  %``X(3872) is not a true molecule,''
  Eur.\ Phys.\ J.\ C {\bf 73}, 2351 (2013)
  [arXiv:1212.0648 [hep-ph]].
  %%CITATION = ARXIV:1212.0648;%%


\bibitem{shen} Cheng-Ping Shen, Proceedings of Few Body 21, Chicago, June 2015.
\bibitem{braaten} E.~Braaten, Proceedings of Few Body 21, Chicago, June 2015.
%\cite{Braaten:2014qka}
\bibitem{Braaten:2014qka} 
  E.~Braaten, C.~Langmack and D.~H.~Smith,
  %``Born-Oppenheimer Approximation for the XYZ Mesons,''
  Phys.\ Rev.\ D {\bf 90}, 014044 (2014)
  [arXiv:1402.0438 [hep-ph]].
  %%CITATION = ARXIV:1402.0438;%%
%\cite{Braaten:2013boa}
\bibitem{Braaten:2013boa} 
  E.~Braaten,
  %``How the Z_c(3900) Reveals the Spectra of Quarkonium Hybrid and Tetraquark Mesons,''
  Phys.\ Rev.\ Lett.\  {\bf 111}, 162003 (2013)
  [arXiv:1305.6905 [hep-ph]].
  %%CITATION = ARXIV:1305.6905;%%
\bibitem{Y(2175)} M.\ Ablikim {\it et al.} [BES collaboration], Phys.\ Rev.\ Lett.\ {\bf 100}, 102003 (2008); 

C.P. Shen {\it et al.} [Belle collaboration], Phys.\ Rev.\ {\bf D80}, 031101 (2009).

%\cite{Fariborz:2006ff}
\bibitem{Fariborz:2006ff} 
  A.~H.~Fariborz, R.~Jora and J.~Schechter,
  %``Model for light scalars in QCD,''
  Phys.\ Rev.\ D {\bf 76}, 014011 (2007)
  [hep-ph/0612200].
  %%CITATION = HEP-PH/0612200;%%
%\cite{Black:2000nc}
\bibitem{Black:2000nc} 
  D.~Black, A.~H.~Fariborz and J.~Schechter,
  %``Mechanism for a next-to-lowest lying scalar meson nonet,''
  AIP Conf.\ Proc.\  {\bf 541}, 100 (2000).
%\cite{Dudek:2009qf}
\bibitem{Dudek:2009qf} 
  J.~J.~Dudek, R.~G.~Edwards, M.~J.~Peardon, D.~G.~Richards and C.~E.~Thomas,
  %``Highly excited and exotic meson spectrum from dynamical lattice QCD,''
  Phys.\ Rev.\ Lett.\  {\bf 103}, 262001 (2009)
  [arXiv:0909.0200 [hep-ph]].
  %%CITATION = ARXIV:0909.0200;%%
%\cite{Dudek:2011tt}
\bibitem{Dudek:2011tt} 
  J.~J.~Dudek, R.~G.~Edwards, B.~Joo, M.~J.~Peardon, D.~G.~Richards and C.~E.~Thomas,
  %``Isoscalar meson spectroscopy from lattice QCD,''
  Phys.\ Rev.\ D {\bf 83}, 111502 (2011)
  [arXiv:1102.4299 [hep-lat]].
  %%CITATION = ARXIV:1102.4299;%%

%\cite{Dudek:2012vr}
\bibitem{opportunities} 
  J.~Dudek {\it et al.},
  %``Physics Opportunities with the 12 GeV Upgrade at Jefferson Lab,''
  Eur.\ Phys.\ J.\ A {\bf 48}, 187 (2012)
  [arXiv:1208.1244 [hep-ex]].
  %%CITATION = ARXIV:1208.1244;%%
%\cite{Dudek:2014qha}
\bibitem{Dudek:2014qha} 
  J.~J.~Dudek {\it et al.} [Hadron Spectrum Collaboration],
  %``Resonances in coupled $\pi K -\eta K$ scattering from quantum chromodynamics,''
  Phys.\ Rev.\ Lett.\  {\bf 113}, 182001 (2014)
  [arXiv:1406.4158 [hep-ph]].
  %%CITATION = ARXIV:1406.4158;%%
%\cite{Wilson:2014cna}
\bibitem{Wilson:2014cna} 
  D.~J.~Wilson, J.~J.~Dudek, R.~G.~Edwards and C.~E.~Thomas,
  %``Resonances in coupled $\pi K, \eta K$ scattering from lattice QCD,''
  Phys.\ Rev.\ D {\bf 91}, 054008 (2015)
  [arXiv:1411.2004 [hep-ph]].
  %%CITATION = ARXIV:1411.2004;%%

%\cite{Shultz:2015pfa}
\bibitem{Shultz:2015pf} 
  C.~J.~Shultz, J.~J.~Dudek and R.~G.~Edwards,
  %``Excited meson radiative transitions from lattice QCD using variationally optimized operators,''
  Phys.\ Rev.\ D {\bf 91}, 114501 (2015)
  [arXiv:1501.07457 [hep-lat]].
  %%CITATION = ARXIV:1501.07457;%%

%\cite{Briceno:2015dca}
\bibitem{Briceno:2015dca} 
  R.~A.~Briceno, J.~J.~Dudek, R.~G.~Edwards, C.~J.~Shultz, C.~E.~Thomas and D.~J.~Wilson,
  %``The resonant $\pi^+\gamma\to\pi^+\pi^0$ amplitude from Quantum Chromodynamics,''
  arXiv:1507.06622 [hep-ph].
  %%CITATION = ARXIV:1507.06622;%%
%\cite{Briceno:2014uqa}

\bibitem{Alde:1988bv} 
  D.~Alde {\it et al.} [IHEP-Brussels-Los Alamos-Annecy(LAPP) Collaboration],
  %``Evidence for a 1-+ Exotic Meson,''
  Phys.\ Lett.\ B {\bf 205}, 397 (1988).
  %%CITATION = PHLTA,B205,397;%%
%\cite{Alde:1988mi}
\bibitem{Alde:1988mi} 
  D.~Alde {\it et al.} [IFVE-Brussels-Annecy-Los-Alamos Collaboration],
  %``OBSERVATION OF AN EXOTIC MESON OF 1.9-GeV MASS DECAYING INTO eta-prime eta,''
  Sov.\ J.\ Nucl.\ Phys.\  {\bf 48}, 1035 (1988)
  [Yad.\ Fiz.\  {\bf 48}, 1727 (1988)].
  %%CITATION = SJNCA,48,1035;%%

\bibitem{chung} S-U.~Chung {\it et al.} [BNL-E852], Phys.\ Rev.\ D{\bf 60}, 092001 (1999) [hep-ex/9902003].
%\cite{Adams:2000zg}
\bibitem{Adams:2000zg} 
 G.~S.~Adams {\it et al.},
  %``Experimental evidence for hadroproduction of exotic mesons,''
  Nucl.\ Phys.\ A {\bf 680}, 335 (2000).
  %%CITATION = NUPHA,A680,335;%%

%\cite{Beladidze:1993km}
\bibitem{Beladidze:1993km} 
  G.~M.~Beladidze {\it et al.} [VES Collaboration],
  %``Study of pi- N ---> eta pi- N and pi- N ---> eta-prime pi- N reactions at 37-GeV/c,''
  Phys.\ Lett.\ B {\bf 313}, 276 (1993).
  %%CITATION = PHLTA,B313,276;%%



%\cite{Zaitsev:2000rc}
\bibitem{Zaitsev:2000rc} 
  A.~Zaitsev [VES Collaboration],
  %``Study of exotic resonances in diffractive reactions,''
  Nucl.\ Phys.\ A {\bf 675}, 155C (2000).
  %%CITATION = NUPHA,A675,155C;%%
%\cite{Krinner:2014mla}
\bibitem{Krinner:2014mla} 
  F.~Krinner [COMPASS Collaboration],
  %``Highlights from COMPASS in hadron spectroscopy,''
  EPJ Web Conf.\  {\bf 96}, 01021 (2015)
  [arXiv:1501.00133 [hep-ex]].
  %%CITATION = ARXIV:1501.00133;%%
%\cite{Dzierba:2005jg}
\bibitem{Dzierba:2005jg} 
  A.~R.~Dzierba {\it et al.},
  %``A Partial wave analysis of the pi- pi- pi+ and pi- pi0 pi0 systems and the search for a J**PC = 1-+ meson,''
  Phys.\ Rev.\ D {\bf 73}, 072001 (2006)
  [hep-ex/0510068].
  %%CITATION = HEP-EX/0510068;%%
%\cite{Alekseev:2009aa}
\bibitem{Alekseev:2009aa} 
  M.~Alekseev {\it et al.} [COMPASS Collaboration],
  %``Observation of a J**PC = 1-+ exotic resonance in diffractive dissociation of 190-GeV/c pi- into pi- pi- pi+,''
  Phys.\ Rev.\ Lett.\  {\bf 104}, 241803 (2010)
  [arXiv:0910.5842 [hep-ex]].
  %%CITATION = ARXIV:0910.5842;%%


%\cite{Adolph:2015pws}
\bibitem{Adolph:2015pws} 
  C.~Adolph {\it et al.} [COMPASS Collaboration],
  %``Observation of a new narrow axial-vector meson $a_1(1420)$,''
  arXiv:1501.05732 [hep-ex].
  %%CITATION = ARXIV:1501.05732;%%
%\cite{Basdevant:2015wma}
\bibitem{Basdevant:2015wma} 
  J.~L.~Basdevant and E.~L.~Berger,
  %``Peak locations and relative phase of different decay modes of the $a_1$ axial vector resonance in diffractive production,''
  Phys.\ Rev.\ Lett.\  {\bf 114}, 192001 (2015)
  [arXiv:1504.05955].

\bibitem{haas} Florian Haas, \lq\lq Two dimensional partial wave analysis of exclusive 190 GeV $\pi^-p$ scattering into the $\pi^-\pi^-\pi^+$ final state at COMPASS'', Ph.D. thesis submitted to Technische Universit\"at M\"unchen, CERN-Thesis-2013-277.
\bibitem{jpac} See https://jpac.jlab.org
%\cite{Deck:1964hm}
\bibitem{Deck:1964hm} 
  R.~T.~Deck,
  %``Kinematical interpretation of the first pi - rho resonance,''
  Phys.\ Rev.\ Lett.\  {\bf 13}, 169 (1964).
  %%CITATION = PRLTA,13,169;%%

%\cite{Bowler:1975my}
\bibitem{Bowler:1975my} 
  M.~G.~Bowler, M.~A.~V.~Game, I.~J.~R.~Aitchison and J.~B.~Dainton,
  %``Diffraction Dissociation, the Deck Mechanism and Diffractive Resonance Production,''
  Nucl.\ Phys.\ B {\bf 97}, 227 (1975).
  %%CITATION = NUPHA,B97,227;%%
\bibitem{Basdevant:1977ya} 
  J.~L.~Basdevant and E.~L.~Berger,
  %``Unitary Coupled-Channel Analysis of Diffractive Production of the a1 Resonance,''
  Phys.\ Rev.\ D {\bf 16}, 657 (1977).
  %%CITATION = PHRVA,D16,657;%%

 %\cite{Khuri:1960zz}
\bibitem{Khuri:1960zz} 
  N.~N.~Khuri and S.~B.~Treiman,
  %``Pion-Pion Scattering and K + /- --> 3pi Decay,''
  Phys.\ Rev.\  {\bf 119}, 1115 (1960).
  %%CITATION = PHRVA,119,1115;%%
%\cite{Danilkin:2014cra}
\bibitem{Danilkin:2014cra} 
  I.~V.~Danilkin, C.~Fernández-Ramírez, P.~Guo, V.~Mathieu, D.~Schott, M.~Shi and A.~P.~Szczepaniak,
  %``Dispersive analysis of ω/ϕ→3π,πγ*,''
  Phys.\ Rev.\ D {\bf 91}, 094029 (2015)
  [arXiv:1409.7708 [hep-ph]].
  %%CITATION = ARXIV:1409.7708;%%
\bibitem{Guo:2014vya} 
  P.~Guo, I.~V.~Danilkin and A.~P.~Szczepaniak,
  %``Dispersive approaches for three-particle final states interaction,''
  arXiv:1409.8652 [hep-ph].
  %%CITATION = ARXIV:1409.8652;%%
%\cite{Guo:2015zqa}
\bibitem{Guo:2015zqa} 
  P.~Guo, I.~V.~Danilkin, D.~Schott, C.~Fernández-Ramírez, V.~Mathieu and A.~P.~Szczepaniak,
  %``Three-body Final State Interaction in $\eta \to 3 \pi$,''
  arXiv:1505.01715 [hep-ph].
  %%CITATION = ARXIV:1505.01715;%%







\end{thebibliography}

\end{document}